\documentclass{elsarticle}
\usepackage{graphicx,color}
\usepackage{amsmath}
\usepackage{ulem}

\usepackage{amsmath}
\usepackage{amssymb}

\def\simle{\mathrel{\rlap{\raise 0.511ex \hbox{$<$}}{\lower 0.511ex \hbox{$\sim$}}}}
\def\simge{\mathrel{ \rlap{\raise 0.511ex \hbox{$>$}}{\lower 0.511ex \hbox{$\sim$}}}}

\newcommand \beq{\begin{eqnarray}}
\newcommand \eeq{\end{eqnarray}} 
\def\simle{\mathrel{\rlap{\raise 0.511ex \hbox{$<$}}{\lower 0.511ex 
\hbox{$\sim$}}}}
\def\simge{\mathrel{ \rlap{\raise 0.511ex 
\hbox{$>$}}{\lower 0.511ex \hbox{$\sim$}}}}

\def\x{{\boldsymbol x}}

\begin{document}
\begin{frontmatter}
\title{Transverse Momentum of Protons, Pions and Kaons in High Multiplicity pp and pA Collisions: Evidence for the Color Glass Condensate?}

\author[bnl,rbrc,ccnu]{Larry McLerran}
\author[krakow]{Michal Praszalowicz}
\author[bnl]{Bj\"orn Schenke}

\address[bnl]{Physics Dept., Bldg. 510A, Brookhaven National Laboratory, Upton, NY 11973, USA}
\address[rbrc]{RIKEN BNL Research Center, Bldg. 510A\\ Brookhaven National Laboratory, Upton, NY 11973, USA}
\address[ccnu]{Physics Department, China Central Normal University, Wuhan, 430079, China}
\address[krakow]{M. Smoluchowski Institute of Physics, Jagellonian University\\ Reymonta 4, 30-059, Krakow, Poland}   

\begin{abstract}
The CMS experiment at the Large Hadron Collider has recently presented data on the average transverse momentum of protons, kaons and pions as a function
of particle multiplicity.  We relate the particle multiplicity to saturation momentum using recently published computations of the interaction radius
determined from the theory of the Color Glass Condensate.  We show that the pp and the pA experimental data scale in terms of these saturation momenta.  
Computing transverse momentum spectra for identified particles using Boltzmann-type distributions and relating different associated
multiplicities using geometric scaling, these simple  distributions reproduce the observed dependence of the mean transverse momentum on particle multiplicities 
seen in both pp and pA interactions for pions to good accuracy, and to fair agreement for protons and kaons.
\end{abstract}
\end{frontmatter}

\section{Introduction}

At very high energies,  produced particle multiplicities can become very large.  The gluons which are ultimately responsible for producing particles
acquire very high densities per unit area, and are controlled by a scale, the saturation 
momentum \cite{Kharzeev:2004if},
\begin{equation}
  Q_{\rm sat}^2 \sim \frac{dN}{dy}\frac{1}{S_{\rm T}}
  \label{Qsatmult}
\end{equation}
where $S_{\rm T}$ is an appropriately defined measure of transverse interaction
area that we discuss in some detail in Sect.~\ref{IA}.
The constant of proportionality in Eq.~(\ref{Qsatmult}) goes as $1/\alpha_{\rm s}$, 
where $\alpha_{\rm s}$ 
is measured at the saturation scale.
For large enough saturation momenta, this dependence is rather weak, and for the analysis in this paper, we will ignore it.

In the theory of the Color Glass Condensate (CGC) 
 \cite{Gribov:1984tu}\nocite{Mueller:1985wy,McLerran:1993ka,McLerran:1993ni}--\cite{Kharzeev:2000ph}, which is supposed to describe the high energy limit of QCD, if we ignore the weak saturation momentum dependence of coupling constant, particle densities for particle species $i$ in proton-proton collisions should scale as \cite{Stasto:2000er,McLerran:2010ex,Praszalowicz:2011rm}
\begin{equation}
\label{eq:eqn1}
  {1 \over S_{\rm T}} { {dN_i} \over {dyd^2p_{\rm T}}} = 
  F_i \left( {p_{\rm T} \over Q_{\rm s}}, {m_i \over Q_{\rm s}}\right) .
 \end{equation}

The form of Eq.\,(\ref{eq:eqn1}) follows from hypothesis of geometric scaling, which states that
differential distributions of charged particles produced in hadronic collisions, which in principle 
should depend on two kinematical variables $p_{\rm T}$ and $s$, depend only on a specific combination 
of them, called scaling variable $\tau=p_{\rm T}/Q_{\rm s}(p_{\rm T}/\sqrt{s})$. Here $Q_{\rm s}$
is a saturation scale introduced in Ref.~\cite{Stasto:2000er}. It should be distinguished from 
$Q_{\rm sat}$ used in Eq.~(\ref{Qsatmult}) which is the scale appearing after integrating differential
distributions, like the one in Eq.~(\ref{eq:eqn1}), over some region of  $p_{\rm T}$. In this case
$Q_{\rm sat}$ should be interpreted as a typical transverse momentum for given energy, atomic number $A$, centrality class, {\it etc.}, and can be though of as a solution of an equation 
$Q_{\rm sat}=Q_{\rm s}(Q_{\rm sat}/\sqrt{s})$ in some particular kinematical region.
Over the limited transverse momentum range needed to determine the average transverse momentum for a particle species,
the $x$ dependence of the saturation momentum is very weak, so in what follows, we ignore it, and the saturation momentum may be thought of as an average saturation momentum scale
appropriate for a typical transverse momentum scale.
Equation (\ref{eq:eqn1}) does a good job \cite{McLerran:2010ex,Praszalowicz:2011rm}
describing the energy dependence of total charged particle production at the Large Hadron Collider (LHC)
\cite{Aamodt:2009aa}\nocite{Aamodt:2010pp,Aamodt:2010ft,Khachatryan:2010xs,Khachatryan:2010us}--\cite{Aad:2010rd}.  In such a circumstance, the majority of particles are pions, and the mass may be ignored.
If, however, we want to study identified particles of species $i$
\cite{Chatrchyan:2012qb}\nocite{CMSpPb,sikler1}--\cite{preghenella},
then the dependence on their masses $m_i$
appears naturally in Eq.~(\ref{eq:eqn1}).  Identified particle spectra are described,
{\it e.g.} by the CMS collaboration \cite{Chatrchyan:2012qb,CMSpPb}, by Tsallis distributions with parameters $n$ and $T$ that are different for pions, kaons and protons. Consequently functions $F_i$ in Eq.~(\ref{eq:eqn1}) depend on particle species. This is further demonstrated in Sect.~\ref{mTT}
where we define effective temperature and the pertinent saturation scale that depend on particle species $i$.

In our analysis, we shall be considering both pp \cite{Chatrchyan:2012qb} and pA \cite{CMSpPb}
collisions using data from the CMS collaboration at the LHC.  In this case, there are two 
saturation momentum scales: that of the proton, $Q_{\rm s}^{({\rm p})}=Q_{\rm p}$, and that of the nucleus, $Q_{\rm s}^{(A)}=Q_{A}$, \cite{Kovchegov:1998bi,Dumitru:2001ux}. 
In this case Eq.\,(\ref{eq:eqn1}) has the additional dependence
\begin{equation}
\label{eqn2}
{ 1\over S_{\rm T}}  {{dN_i} \over {dyd^2p_{\rm T}}} = 
F_i \left( {p_{\rm T}\over Q_{\rm p}},{ m_i \over Q_{\rm p}}, {Q_{\rm p} \over {Q_A}} \right).
\end{equation}
For low multiplicities, the proton saturation momentum scale $Q_{\rm p}$ is less than that of the nucleus, $Q_A$, but at high energies and/or central rapidities, for high multiplicity fluctuations, they should eventually become equal, $Q_{\rm p}/Q_A = 1$ as in pp collisions. 
The equality for high multiplicities follows because once the proton saturation momentum  equals that of the nucleus, then there is no gap in the transverse momentum spectrum between that of
the saturation momentum of the proton and that of the nucleus.  This occurs by the saturation momentum of the proton increasing without much change in that of the nucleus.  If we go to higher density, then we might expect the saturation momentum of the proton and that of the nucleus to increase in tandem, since this would require making a rated fluctuation needed to generate an asymmetric distribution.   

When we apply Eq.\,(\ref{eqn2}) to pA collisions, 
we shall ignore the dependence on $Q_{\rm p}/Q_A$, assuming the dependence is weak, an assumption justifiable for very high multiplicities.

An issue not tested by such a geometric scaling description is whether this scaling works as a function of multiplicity and for identified particle species.
There has been a recent analysis by the CMS collaboration, that geometric scaling does a fair job of describing pion, kaon and proton production in proton-proton collisions \cite{sikler2}. 
In this paper we explore how well geometric scaling describes the now available LHC data.  In particular, we will attempt to describe the recently presented CMS data on the mean transverse momentum of produced particles as a function of particle multiplicity. 

The paper is organized as follows. In Sect.~\ref{IA} we discuss the emergence of the transverse interaction 
area, $S_{\rm T}$, introduced in Eq.~(\ref{Qsatmult}) which is an essential novel ingredient of our 
analysis. We show next in Sect.~\ref{scaling} that data for mean transverse momentum as function of
the square root of multiplicity (more precisely of $\sqrt{N_{\rm track}}\,$) scale with $\sqrt{S_{\rm T}}$. 
We do this by comparing
data for pp and pA collisions. In Sect.~\ref{mTT} we show that simple thermal distributions
explain geometric scaling observed in the data. Finally in Sect.~\ref{concl} we present conclusions.

\section{The Interaction Area}
\label{IA}

In order to compute the saturation momentum, one needs the ratio of particle multiplicity to interaction area.  The particle multiplicity
can be taken as an input.  To compute the interaction area as a function of centrality for both pp and pA collisions, we use the result of the computation of Bzdak et. al. \cite{Bzdak:2013zma} in the IP-Glasma model \cite{Schenke:2012wb,Schenke:2012hg}.  This is a computation based on an impact parameter description of pp collisions, combined with an underlying description of particle production based on the theory of the Color Glass Condensate.  It has several remarkable features.  First, the interaction radius is approximately a linear function
of $(dN/dy)^{1/3}$ for $dN/dy$ less than some critical value.  Eventually the radius saturates to a constant that is smaller for pp than pA collisions.  Until the radius for pp collisions saturates, the radii for pp and pA collisions as a function of multiplicity are nearly the same.  

The saturation of the radius $R$ as a function of $(dN/dy)^{1/3}$ can be understood in the following way. 
For smaller multiplicities, the number of produced particles is proportional to the interaction volume $\propto R^3$.  

Once maximal overlap ($b=0\,{\rm fm}$) is achieved, higher multiplicities can only be reached by certain color charge fluctuations, 
which do not increase the size of the system. This argument also holds for pA collisions, where $b=0\,{\rm fm}$ corresponds to an overlap of the proton with the densest region of the nucleus. 

The greatest uncertainty in this computation is for the smallest multiplicity collisions.  We will take the minimum multiplicity where such a computation may be reliable to be
the minimum bias multiplicity in pp collisions, which we will take to be around 5 charged particles per unit rapidity.  In addition, the ratio of the radius for pp and pA can be more or less unambiguously defined and is largely independent of a cutoff in the energy density at which it is measured, the absolute values of the radii depend more strongly on the precise definition. This leads to an overall constant uncertainty in the saturation momentum, which will however not be important in our scaling analysis.

\begin{figure}[h]
\begin{center}
\includegraphics[width=9cm]{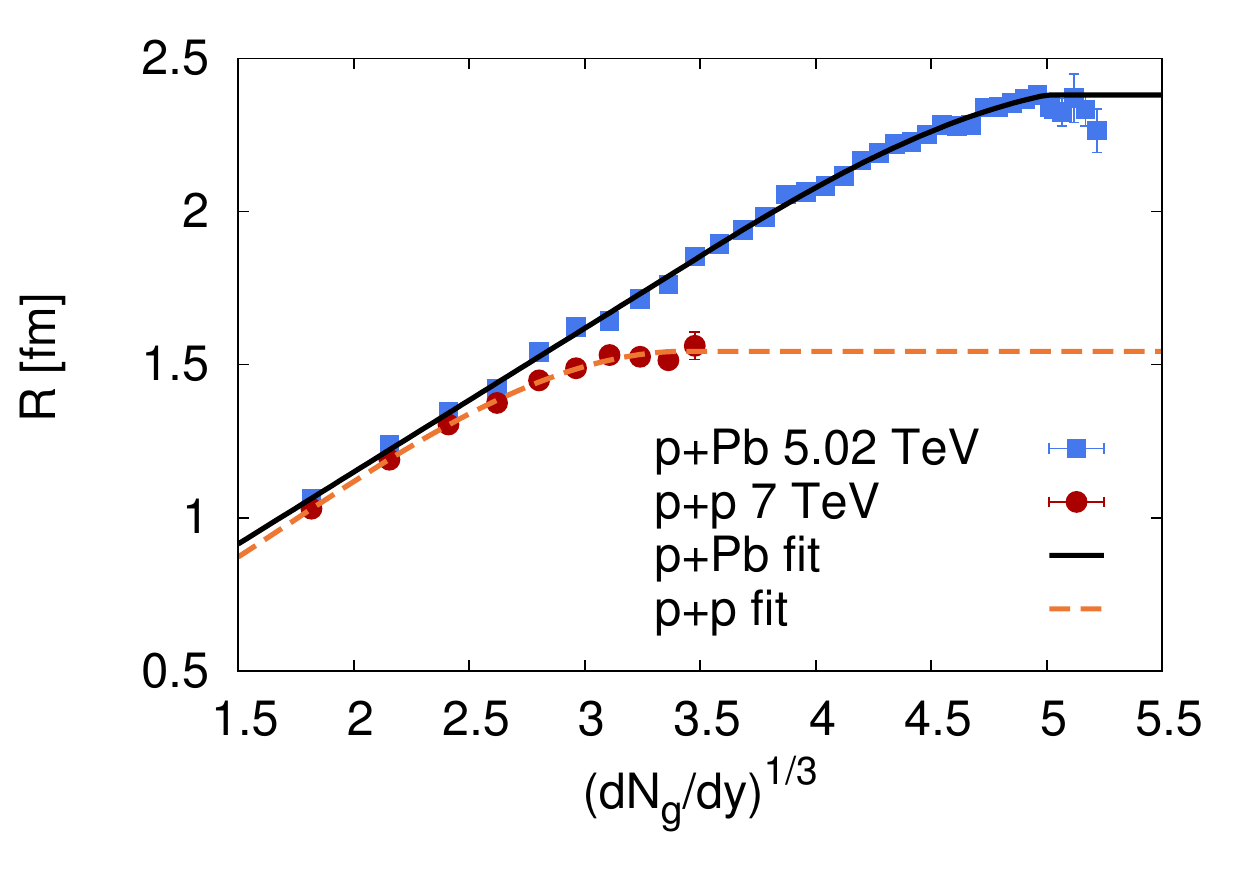}
\caption{\label{multi}Radius $R_{\rm pPb}=\sqrt{S_{\rm  pPb}/\pi}$ for pPb collisions and $R_{\rm pp}=\sqrt{S_{\rm pp}/\pi}$ for 
pp collisions vs $(dN_g/dy)^{1/3}$ as computed in the IP-Glasma model \cite{Bzdak:2013zma} together with the corresponding fits. \label{fig:bzdak}}
\end{center} 
\end{figure}

The results for the radii of \cite{Bzdak:2013zma} are shown in Fig.\,\ref{fig:bzdak},
where the fitted curves are our parametrization of the computations. $R_{\rm pPb}$ as a function of the gluon multiplicity 
is parametrized as
\begin{equation}
  R_{\rm pPb} = 1\,{\rm fm}\times f_{\rm pPb}\left(\sqrt[3]{dN_g/dy}\right)
\end{equation}
with
\begin{equation}
  f_{\rm pPb}(x) = \left\{ \begin{array}{ll}
         0.21 + 0.47\,x  & \mbox{if $x < 3.5$,}\\
         1.184 - 0.483\,x  + 0.305\,x^{2}  - 0.032\,x^3 & \mbox{if $3.5 \leq x < 5 $,}\\
         2.394 & \mbox{if $x \geq 5 $.}
       \end{array} \right. 
\end{equation}
The gluon multiplicity $dN_g/dy$ can be approximately related to the number of tracks seen in the CMS experiment by
\begin{equation}
  {{dN_g} \over {dy}} \approx {3 \over 2} {1 \over {\Delta \eta}} N_{\rm track}
\end{equation}
where $\Delta \eta \sim 4.8$ units of pseudo-rapidity. 
The cutoff in this formula, where we no longer trust the computation of the radius, corresponds to $N_{\rm track} \sim 20$.
For the pp radius we take
\begin{equation}
 R_{\rm pp} = 1\,{\rm fm}\times f_{\rm pp}\left(\sqrt[3]{dN_g/dy}\right)
\end{equation}
with
\begin{equation}
 f_{\rm pp}(x) = \left\{ \begin{array}{ll}
         0.387 + 0.0335 x + 0.274\,x^2 - 0.0542\,x^3  & \mbox{if $x < 3.4$,}\\
         1.538 & \mbox{if $x \geq 3.4 $.}
       \end{array} \right. 
\end{equation}
 
 \section{Scaling of the Measured $\langle p_{\rm T}\rangle$ as a Function of Multiplicity}
 \label{scaling}
 
\begin{figure}[h]
\begin{center}
\includegraphics[width=10cm]{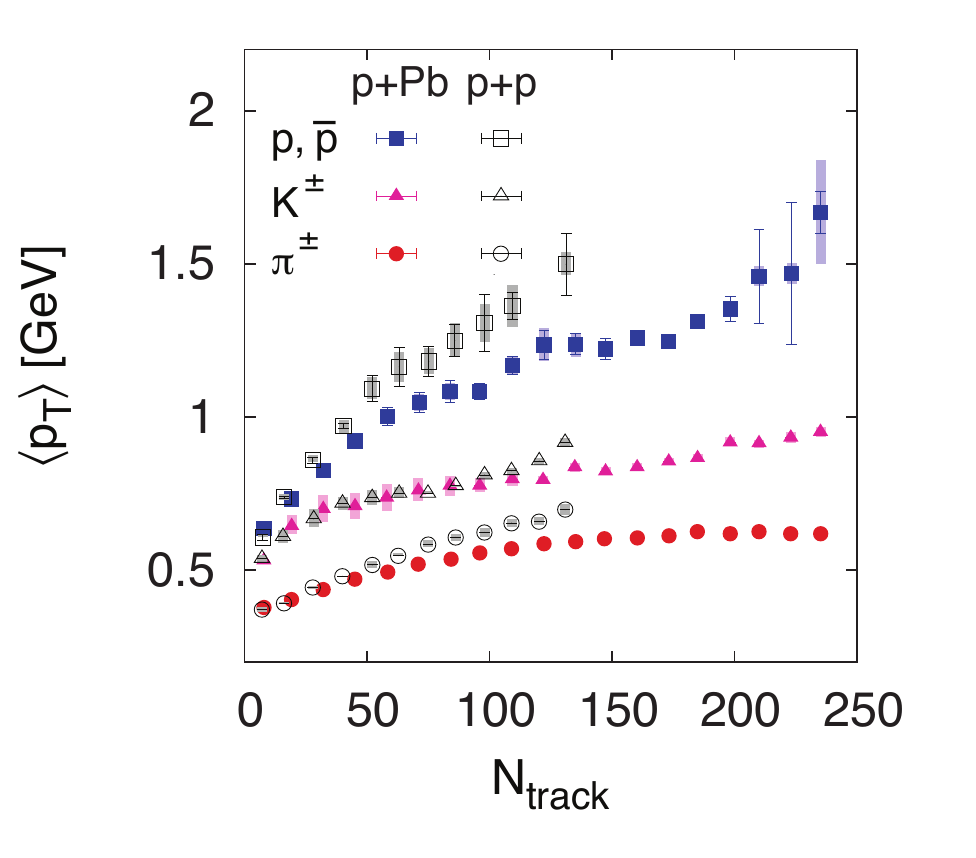}
\caption{ Mean transverse momentum in pp and pPb collisions measured by the CMS collaboration vs. $N_{\rm track}$. \label{fig:CMSData}}
\end{center} 
\end{figure}

\begin{figure}[h]
\begin{center}
\includegraphics[width=10cm]{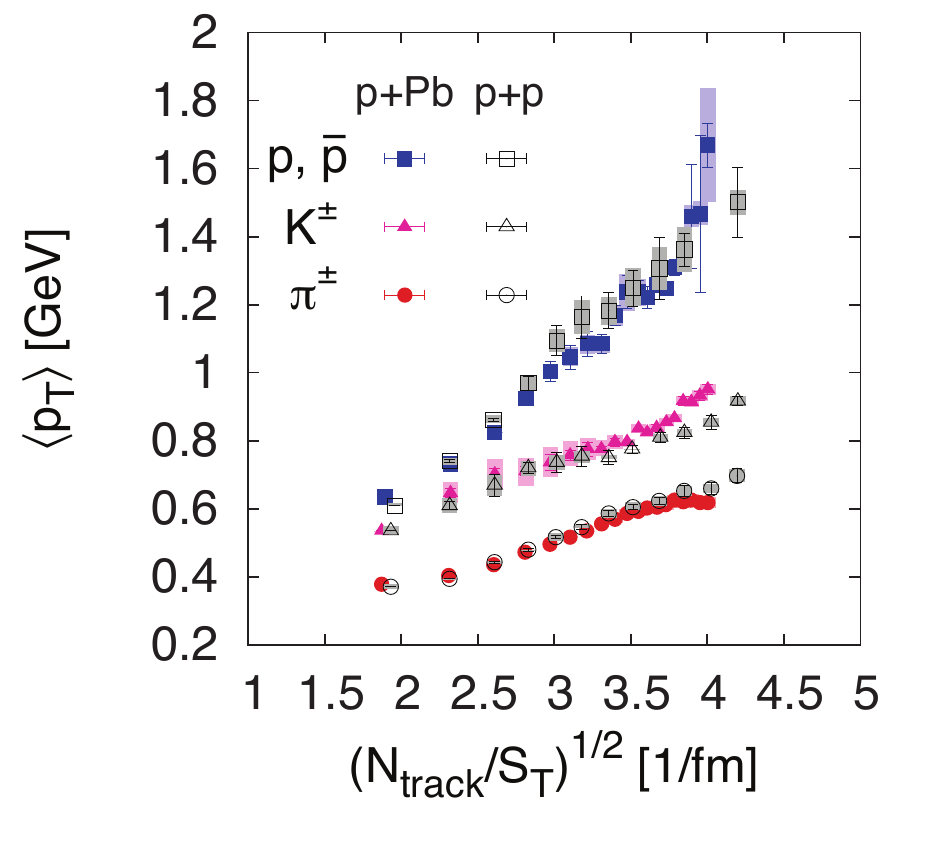}
\caption{ Mean transverse momentum in pp and pPb collisions measured by the CMS collaboration vs$.$ $\sqrt{N_{\rm track}}$ scaled by the square root of the transverse area $S_{\rm pp}=\pi R_{\rm pp}^2$ and  $S_{\rm pPb}=\pi R_{\rm pPb}^2$, respectively. \label{fig:scaledCMSData}
}
\end{center} 
\end{figure}

 If particle distributions have the scaling property of Eq.\,(\ref{eq:eqn1}), then for a single particle species, 
 the distributions will map into one another if plotted vs. the saturation momentum. The saturation momentum squared would be linear in the associated multiplicity were it not for the dependence of area of the interaction region upon multiplicity.  The data presented by CMS is shown in Fig.\,\ref{fig:CMSData}.  We have then computed the saturation momentum associated with each multiplicity for pp and pA, then
replotted the data as a function of the corresponding saturation momenta, or the square root of the ratio of multiplicity and the transverse area.  We see that to within experimental accuracy, the identified proton, kaon and pion spectra map into one another as shown in Fig.\,\ref{fig:scaledCMSData}. 
A physical interpretation of this behavior is that local particle production is determined by the system's properties within flux tubes of size $1/Q_{\rm sat}$ and not affected by the total system size.

 \section{Do the Identified Particle $p_{\rm T}$ Distributions Obey Geometric Scaling?}
 \label{mTT}
 


We describe the transverse momentum spectra of particle species $i$ by the distribution 
\begin{equation}
  f(m_i,T^{\rm eff}_i,p_{\rm T}) \sim e^{-(m_T)_i/T_i^{\rm eff}}\,, 
\end{equation}
where $(m_T)_i = \sqrt{p_{\rm T}^2+m_i^2}$ and 
the effective temperature $T^{\rm eff}$ is parametrized as 
\begin{equation}\label{eq:T}
  T^{\rm eff}_i = \kappa_i\, Q_{\rm p} = \kappa_i\, \sqrt{\frac{N_{\rm track}}{S_{\rm T}}}\,,
\end{equation}
where we have neglected any dependence on $Q_A$ in pA collisions as discussed above.
From this distribution we can compute the 
dependence of the mean transverse momentum of particle species $i$ on multiplicity
using
\begin{equation}
  \langle p_{\rm T} \rangle_i = \frac{\int p_{\rm T}^2 dp_{\rm T} \, f(m_i,T^{\rm eff}_i,p_{\rm T})}{\int p_{\rm T} dp_{\rm T}\, f(m_i,T^{\rm eff}_i,p_{\rm T})} 
  = \frac{m_i^2}{m_i+T^{\rm eff}_i} K_2\left(\frac{m_i}{T^{\rm eff}_i}\right) e^{m_i/T^{\rm eff}_i}\,,
\end{equation}
where $K_2$ is the modified Bessel function of the second kind.

\begin{figure}[htb]
  \begin{center}
    \includegraphics[width=10cm]{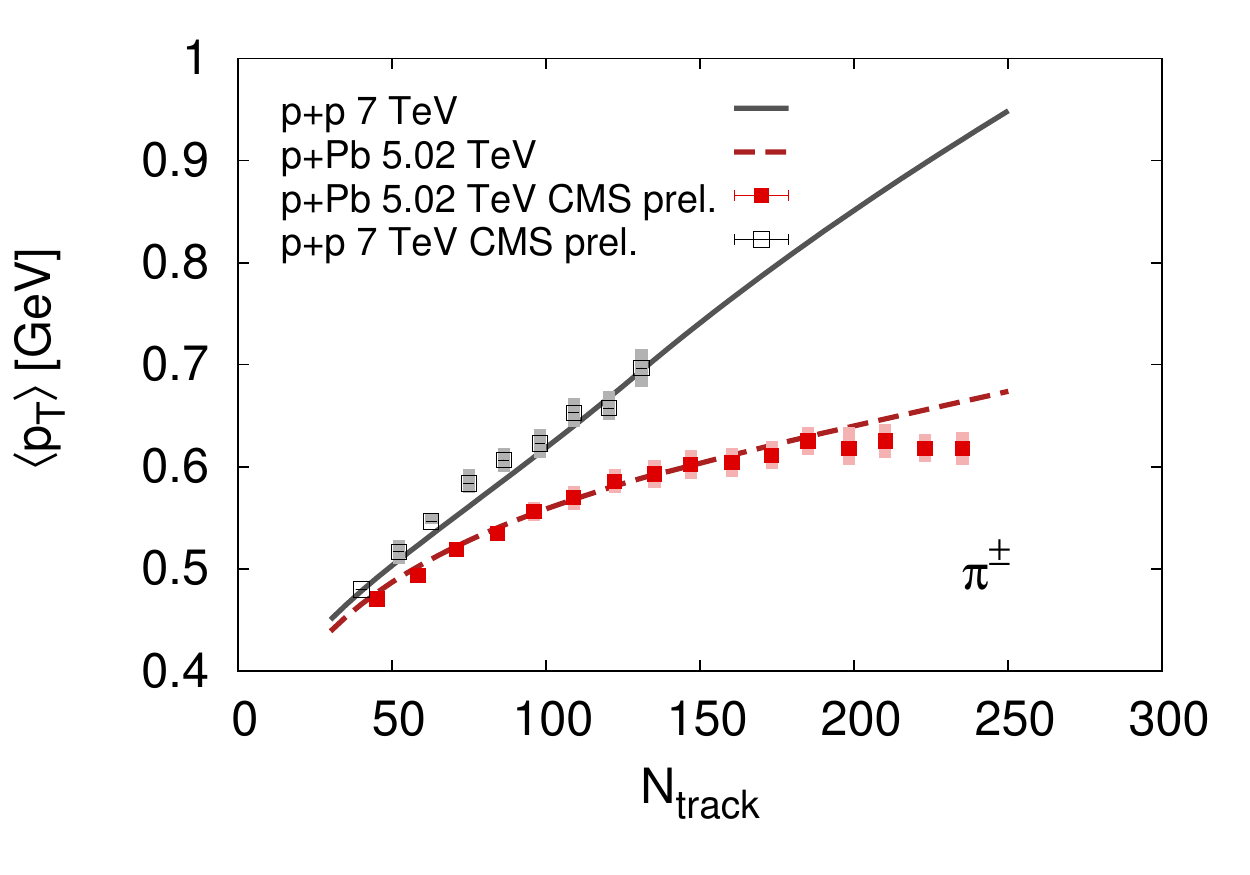}
    \caption{Mean transverse momentum in pp and pPb collisions for pions vs. $N_{\rm track}$ compared to experimental data from the CMS collaboration. The difference between pp and pPb results stems 
    solely from different transverse areas.  \label{fig:meanptpi}
 }
  \end{center} 
\end{figure}

\begin{figure}[htb]
  \begin{center}
    \includegraphics[width=10cm]{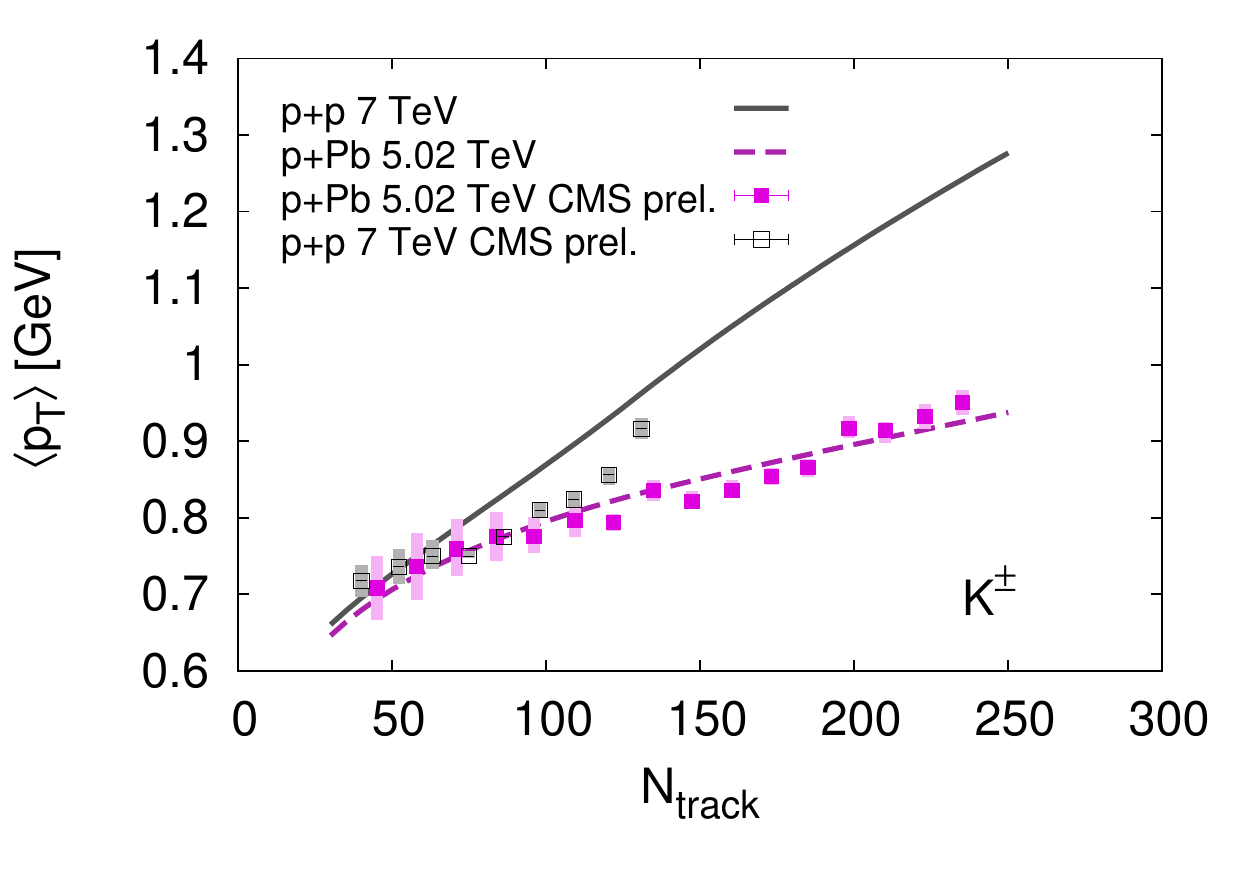}
    \caption{Mean transverse momentum in pp and pPb collisions for kaons vs. $N_{\rm track}$ compared to experimental data from the CMS collaboration. The difference between pp and pPb results stems 
    solely from different transverse areas.   \label{fig:meanptK}
}
  \end{center} 
\end{figure}

\begin{figure}[htb]
  \begin{center}
    \includegraphics[width=10cm]{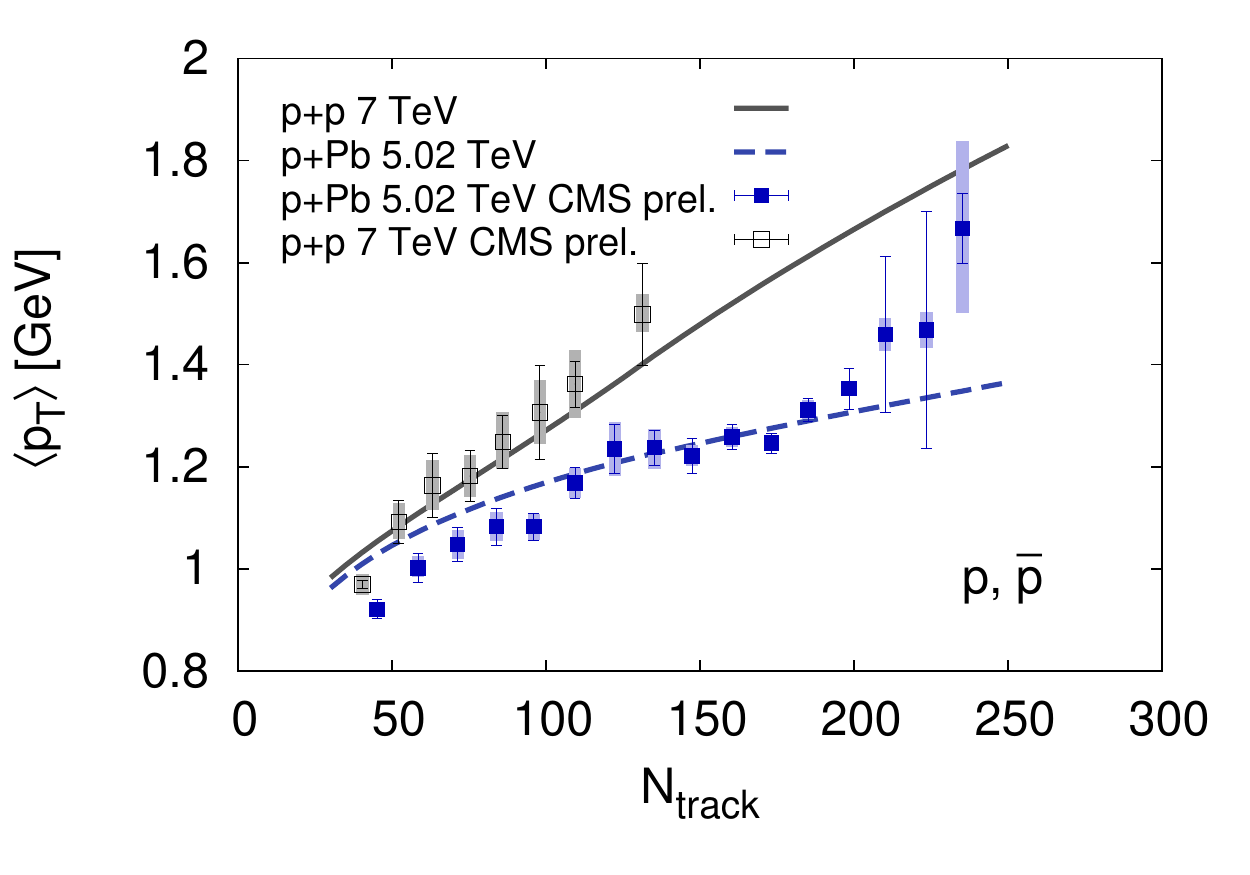}
    \caption{Mean transverse momentum in pp and pPb collisions for protons vs. $N_{\rm track}$ compared to experimental data from the CMS collaboration. The difference between pp and pPb results stems 
    solely from different transverse areas.   \label{fig:meanptp}
}
  \end{center} 
\end{figure}

\begin{figure}[h]
  \begin{center}
    \includegraphics[width=10cm]{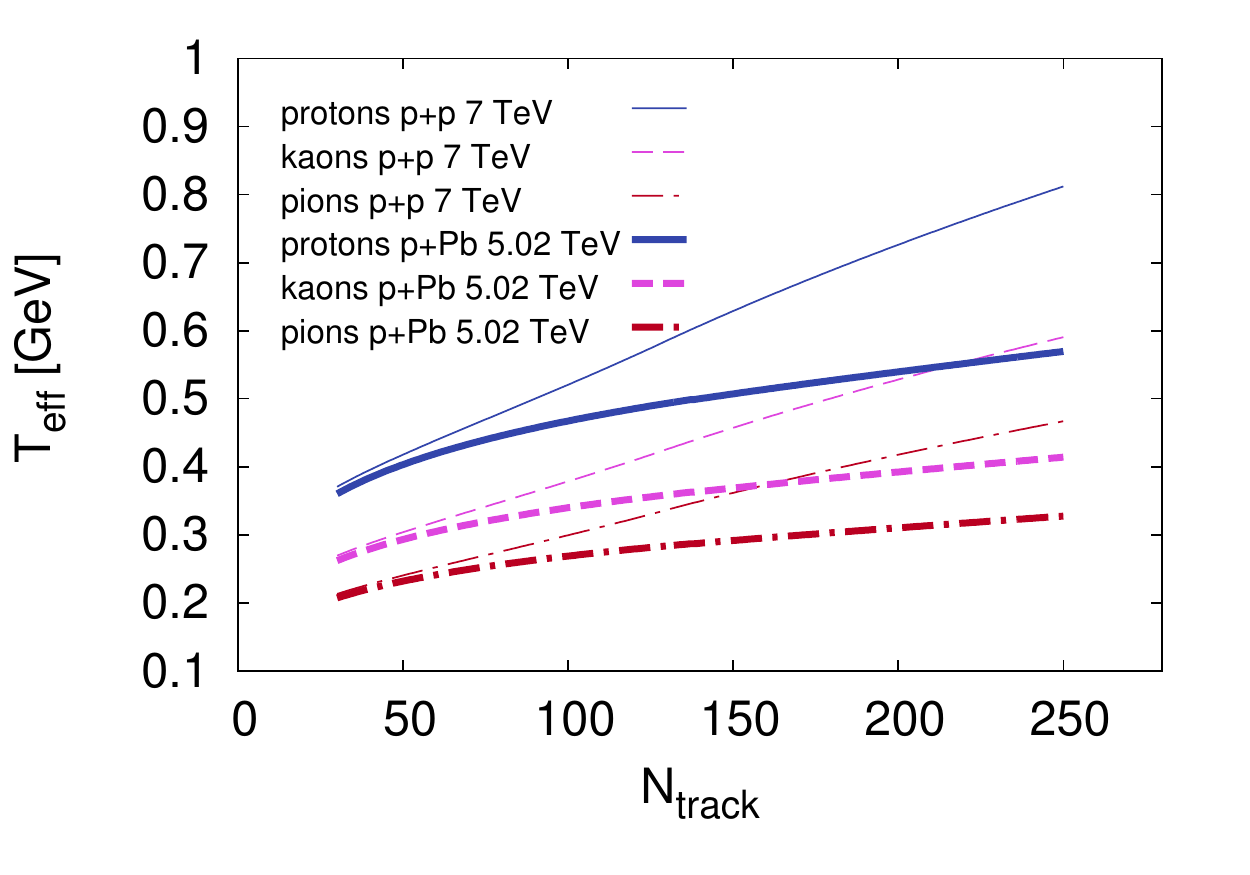}
    \caption{ Effective temperature as a function of $N_{\rm track}$ in pp (thin lines) and pPb (thick lines) collisions for protons, kaons 
      and pions. \label{fig:T}}
  \end{center} 
\end{figure}

Results for protons, kaons, and pions are shown in Figs.\,\ref{fig:meanptpi} - \ref{fig:meanptp}.
The only difference between results for pp and pPb collisions is the transverse size entering 
through $T^{\rm eff}$ according to Eq.\,(\ref{eq:T}).
We fixed the parameter $\kappa_i$ in Eq.\,(\ref{eq:T}) for each particle species $i$ to fit one value of $\langle p_{\rm T} \rangle_i$ in pp collisions.
We chose to do this at a value between $N_{\rm track}=50$ and $60$ for protons and kaons and for $N_{\rm track}\approx 30$ for pions.
The description of the pion $\langle p_{\rm T}\rangle$ is excellent in both pp and pA collisions, while for protons we find some deviations especially at low
multiplicities. For kaons there is disagreement between the pp data and the calculation. Generally, however, qualitative agreement is found for all particle species.  
For completeness, the resulting effective temperatures for the different particle species and different collision systems are shown in Fig.\,\ref{fig:T}.

\section{Conclusions}
\label{concl}

It appears that the geometric scaling \cite{Stasto:2000er} and the specific behavior of transverse
interaction area $S_{\rm T}$ with multiplicity
predicted by the theory of the Color Glass Condensate \cite{Bzdak:2013zma}
 describe very well generic features of the mean
transverse momenta of produced particles as a function of centrality at LHC energies 
for pp and pA collisions.  Once identified particle distributions are available
as a function of multiplicity for both pp and pA collisions for a larger multiplicity range
and possibly for the higher energies, 
it will be very interesting to see how far the scaling hypothesis will go in describing such distributions.
The energy dependence of  minimum bias pp collisions for charged particle distributions, and identified particle distributions seem to provide a good description.

One of the key issues in pA collisions is whether some degree
of collectivity is needed to describe the data \cite{Bozek:2011if}, at least for the highest
multiplicities. In this note we show
that simple scaling arguments can explain the flattening of the mean $p_{\rm T}$
for identified particles with $\sqrt[3]{dN/dy}$ which is often attributed to
flow.  Once new data characterizing pPb
collisions, especially on interferometric radii \cite{Bozek:2013df}, are available, 
then one will be able to
distinguish between different scenarios of particle production mechanisms in pA collisions.

One might ask about the implications of the scaling discussed in this paper
for heavy ion collisions.  In such collisions the collision area is controlled by the geometry of the collisions,
and the typical multiplicity scales with the number of participants.  However, in heavy-ion collisions significant final state interactions of particles should generate flow, and modify the
transverse momentum.  The good description of heavy ion collisions  by hydrodynamic simulations suggests, at least for central collisions, that initial state effects on the typical transverse momentum are largely washed out by final state interactions.

\section*{Acknowledgements}
The research of  L. McLerran and B. Schenke is supported under DOE Contract No. DE-AC02-98CH10886. The research of M. Praszalowicz is supported by the Polish NCN 
grant 2011/01/B/ST2/00492.


\begin{thebibliography}{00}

\bibitem{Kharzeev:2004if}
  D.~Kharzeev, E.~Levin and M.~Nardi,
  Nucl.\ Phys.\ A {\bf 747} (2005) 609
  [hep-ph/0408050].

\bibitem{Gribov:1984tu}
  L.~V.~Gribov, E.~M.~Levin and M.~G.~Ryskin,
  Phys.\ Rept.\  {\bf 100} (1983) 1.

\bibitem{Mueller:1985wy}
  A.~H.~Mueller and J.~-w.~Qiu,
  Nucl.\ Phys.\ B {\bf 268} (1986) 427.

\bibitem{McLerran:1993ka}
  L.~D.~McLerran and R.~Venugopalan,
  Phys.\ Rev.\ D {\bf 49} (1994) 3352
  [hep-ph/9311205].

\bibitem{McLerran:1993ni}
  L.~D.~McLerran and R.~Venugopalan,
  Phys.\ Rev.\ D {\bf 49} (1994) 2233
  [hep-ph/9309289].

\bibitem{Kharzeev:2000ph}
  D.~Kharzeev and M.~Nardi,
  Phys.\ Lett.\ B {\bf 507} (2001) 121
  [nucl-th/0012025].
  
\bibitem{Stasto:2000er}
  A.~M.~Stasto, K.~J.~Golec-Biernat and J.~Kwiecinski,
  Phys.\ Rev.\ Lett.\  {\bf 86} (2001) 596
  [hep-ph/0007192].
  
\bibitem{McLerran:2010ex}
  L.~McLerran and M.~Praszalowicz,
  Acta Phys.\ Polon.\ B {\bf 41} (2010) 1917
  [arXiv:1006.4293 [hep-ph]] and 
  Acta Phys.\ Polon.\ B {\bf 42} (2011) 99
  [arXiv:1011.3403 [hep-ph]].
  
\bibitem{Praszalowicz:2011rm}
  M.~Praszalowicz,
  Acta Phys.\ Polon.\ B {\bf 42} (2011) 1557
  [arXiv:1104.1777 [hep-ph]].
 

\bibitem{Aamodt:2009aa}
  KAamodt {\it et al.}  [ALICE Collaboration],
  Eur.\ Phys.\ J.\ C {\bf 65} (2010) 111
  [arXiv:0911.5430 [hep-ex]].

\bibitem{Aamodt:2010pp}
  K.~Aamodt {\it et al.}  [ALICE Collaboration],
  Eur.\ Phys.\ J.\ C {\bf 68} (2010) 345
  [arXiv:1004.3514 [hep-ex]].
  
\bibitem{Aamodt:2010ft}
  K.~Aamodt {\it et al.}  [ALICE Collaboration],
  Eur.\ Phys.\ J.\ C {\bf 68} (2010) 89
  [arXiv:1004.3034 [hep-ex]].

\bibitem{Khachatryan:2010xs}
  V.~Khachatryan {\it et al.}  [CMS Collaboration],
  JHEP {\bf 1002} (2010) 041
  [arXiv:1002.0621 [hep-ex]].
  
\bibitem{Khachatryan:2010us}
  V.~Khachatryan {\it et al.}  [CMS Collaboration],
  Phys.\ Rev.\ Lett.\  {\bf 105} (2010) 022002
  [arXiv:1005.3299 [hep-ex]].
  
\bibitem{Aad:2010rd}
  G.~Aad {\it et al.}  [ATLAS Collaboration],
  Phys.\ Lett.\ B {\bf 688} (2010) 21
  [arXiv:1003.3124 [hep-ex]].
  
\bibitem{Chatrchyan:2012qb}
  S.~Chatrchyan {\it et al.}  [CMS Collaboration],
  Eur.\ Phys.\ J.\ C {\bf 72} (2012) 2164
  [arXiv:1207.4724 [hep-ex]]
  
 \bibitem{CMSpPb}
 [CMS Collaboration] CERN preprint CMS PAS HIN-12-016.
 
 \bibitem{sikler1}
``Measurements of Hadron Production in pPB Collisions in CMS",
F. Sikler for the CMS collaboration, Workshop on Proton-Nucleus Collisions at the LHC, Trento, IT, 6, May 2013.
 
 \bibitem{preghenella}
``ALICE Results on Identified Particle Spectra in p-Pb, Collisions'', R. Preghenella for the ALICE collaboration,
Workshop on Proton-Nucleus Collisions at the LHC, Trento, IT, 6, May 2013
  
\bibitem{Kovchegov:1998bi}
  Y.~V.~Kovchegov and A.~H.~Mueller,
  Nucl.\ Phys.\ B {\bf 529} (1998) 451
  [hep-ph/9802440].

\bibitem{Dumitru:2001ux}
  A.~Dumitru and L.~D.~McLerran,
  Nucl.\ Phys.\ A {\bf 700} (2002) 492
  [hep-ph/0105268].
  
\bibitem{sikler2}
F. Sikler, private communication.

\bibitem{Bzdak:2013zma}
  A.~Bzdak, B.~Schenke, P.~Tribedy and R.~Venugopalan,
  arXiv:1304.3403 [nucl-th].
  
\bibitem{Schenke:2012wb} 
  B.~Schenke, P.~Tribedy and R.~Venugopalan,
  Phys.\ Rev.\ Lett.\  {\bf 108}, 252301 (2012)
  [arXiv:1202.6646 [nucl-th]].

\bibitem{Schenke:2012hg} 
  B.~Schenke, P.~Tribedy and R.~Venugopalan,
  Phys.\ Rev.\ C {\bf 86}, 034908 (2012)
  [arXiv:1206.6805 [hep-ph]].

\bibitem{Bozek:2011if}
  P.~Bozek,
  Phys.\ Rev.\ C {\bf 85} (2012) 014911
  [arXiv:1112.0915 [hep-ph]].
  
\bibitem{Bozek:2013df}
  P.~Bozek and W.~Broniowski,
  Phys.\ Lett.\ B {\bf 720} (2013) 250
  [arXiv:1301.3314 [nucl-th]].




\end{thebibliography}
\end{document}